\documentstyle[prl,twocolumn,aps]{revtex}

\newcommand{\be}{\begin{equation}}
\newcommand{\ee}{\end{equation}}

\begin{document}
\twocolumn[\hsize\textwidth\columnwidth\hsize\csname @twocolumnfalse\endcsname
\draft
\title{{Teleparallel Equivalent of the Kaluza-Klein Theory}}
\author{V. C. de Andrade, L. C. T. Guillen and J. G. Pereira}
\address{Instituto de F\'{\i}sica Te\'orica \\
Universidade Estadual Paulista \\
Rua Pamplona 145 \\ 01405-900 S\~ao Paulo SP \\ Brazil}
\date{\today}
\maketitle

\begin{abstract}
Relying upon the equivalence between a gauge theory for the translation group and general
relativity, a teleparallel version of the original Kaluza-Klein theory is developed. In
this model, only the internal space (fiber) turns out to be five dimensional, spacetime
being kept always four dimensional. A five-dimensional translational gauge theory is
obtained which unifies, in the sense of Kaluza-Klein theories, gravitational and
electromagnetic interactions.
\end{abstract}


\vskip1pc]

\section{Introduction}

In ordinary Kaluza-Klein theories~\cite{mkkt}, the geometrical approach of general
relativity is adopted as the paradigm for the description of all other interactions of
nature. In the original Kaluza-Klein theory, for example, gravitational and
electromagnetic fields are described by a Hilbert-Einstein Lagrangian in a five-dimensional
spacetime.  On the other hand, it has already been shown that, at least macroscopically,
general relativity is equivalent to a gauge theory for the translation group~\cite{hs}. In
this theory, known as the {\it teleparallel equivalent of general relativity}~\cite{maluf},
the fundamental field is the Cartan connection, a connection presenting torsion, but no
curvature. This equivalence gives rise to new perspectives for the study of unified
theories. In fact, instead of using the geometrical description of general relativity, we
can adopt the gauge description as the basic paradigm, and in this way construct what we
call the {\it teleparallel equivalent of the Kaluza-Klein theory}. According to this approach,
both gravitational and electromagnetic fields turn out to be described by a gauge-type
Lagrangian. Such a construction will be the main purpose of this paper, which is organized
as follows. In Sec. II, the teleparallel version of the Kaluza-Klein model is
introduced, and its main features described. The coupling of a general matter field with
gravitational and electromagnetic fields is studied in Sec. III, and finally, in Sec.
IV, we draw the conclusions of the paper.

\section{Teleparallel Kaluza-Klein Model}

We denote by $x^{\mu} (\mu, \nu, \rho,...  = 0,1,2,3)$ the coordinates of spacetime (base
space), and by $x^{a} (a,b,c,...  = 0,1,2,3)$ the coordinates of the tangent space (fiber),
assumed to be a Minkowski space with the metric
\begin{equation}
\eta_{ab}=\mbox{diag}(+1,-1,-1,-1).
\label{1.1}
\end{equation}
The tangent space indices, therefore, are raised and lowered with the Lorentzian metric
$\eta_{a b}$, while the spacetime indices are raised and lowered with the Riemannian
metric $g_{\mu \nu} = \eta_{a b} h^a{}_{\mu} \, h^b{}_{\nu}$, where
\be
h^{a}{}_{\mu} = \partial_{\mu}x^{a} + c^{-2}A^{a}{}_{\mu}  
\label{2.17}
\ee
is a nontrivial tetrad field, with $c$ the speed of light and $A^a{}_{\mu}$ the
gravitational gauge potential. The presence of a nontrivial tetrad field induces on
spacetime a teleparallel structure which is directly related to the presence of the
gravitational field. In fact, given a nontrivial tetrad $h^{a}{}_{\mu}$, the Cartan
connection
\be
\Gamma^{\rho}{}_{\mu \nu} = h_{a}{}^{\rho}\partial_{\nu}h^{a}{}_{\mu}
\label{2.19}
\ee
can be defined, which is a connection presenting torsion, but no curvature, and which
defines a teleparallel structure in spacetime~\cite{livro}.

In the framework of the teleparallel description of gravitation, the
action describing a particle of mass $m$ and charge $e$ submitted to both an
electromagnetic ($A_{\mu}$) and a gravitational ($A^a{}_{\mu}$) field
is~\cite{landau,paper10}
\begin{equation}
S=\int_{a}^{b}\left[ - m \, c \, d\sigma - \frac{m}{c} \, A^{a}{}_{\mu} \, u_{a} \,
dx^{\mu} - 
\frac{e}{c} \, A_{\mu } \ dx^{\mu }\right] ,  \label{3.1}
\end{equation}
where $d\sigma=(\eta_{a b} dx^a dx^b)^{1/2}$ is the invariant tangent-space interval, and
$u_a = {d x_a}/{d \sigma}$. The corresponding equation of motion is
\begin{equation}
c^{2} h^{a}{}_{\rho } \frac{du_{a}}{ds} = F^{a}{}_{\rho \mu }
u_{a}u^{\mu} + \frac{e}{m} \, F_{\rho \mu } u^{\mu} ,
\label{3.2}
\end{equation}
where $ds=(g_{\mu \nu} dx^\mu dx^\nu)^{1/2}$ is the invariant spacetime interval,
$F^{a}{}_{\mu \nu}$ and $F_{\mu \nu}$ are respectively the gravitational and the
electromagnetic field strengths, and $u^\mu = {d x^\mu}/{d s}$ is the spacetime
four velocity. The gravitational field strength satisfies the relation
$F^{a}{}_{\mu \nu}$ = $c^{2}h^{a}{}_{\rho}T^{\rho}{}_{\mu \nu}$, with
\be
T^{\rho}{}_{\mu \nu} = \Gamma^{\rho}{}_{\nu \mu} - \Gamma^{\rho}{}_{\mu \nu}
\label{2.21a}
\ee
the torsion of the Cartan connection.

Now, inspired by the similarity between the gravitational and
the electromagnetic interactions, we choose the $U(1)$ gauge index of
the electromagnetic theory to be ``5'', which allows us to define a unified gauge
potential by 
\[
{\mathcal A}^{A}{}_{\mu} = \left( A^{a}{}_{\mu }, A^{5}{}_{\mu} \right) , 
\]
where $A^{5}{}_{\mu} = \kappa^{-1} ({e}/{m}) A_\mu$, with $\kappa$ a dimensionless
parameter to be determined later. Accordingly, we can define a unified field strength,
\[
{\mathcal F}^{A}{}_{\mu \nu} = \left( F^{a}{}_{\mu \nu},
F^{5}{}_{\mu \nu} \right), 
\]
with $F^{5}{}_{\mu \nu} = \kappa^{-1} ({e}/{m}) F_{\mu \nu}$. In terms of the
potential
${\mathcal A}^{A}{}_{\mu}$, therefore, the field strength is
\begin{equation}
{\mathcal F}^{A}{}_{\mu \nu } = \partial_{\mu} {\mathcal A}^{A}{}_{\nu } -
\partial_{\nu} {\mathcal A}^{A}{}_{\mu}.  \label{3.7}
\end{equation}

Implicit in the above definitions is the introduction of an internal
five-dimensional space $M^5$, given by the product between the Minkowski space
$M^4$ and the circle $S^1$: $M^5$ = $M^4 \otimes S^1$.
A point in this space is determined by the coordinates
$x^A$ = $(x^a, x^5)$, where $x^a$ are the coordinates of $M^4$, and $x^5 = \kappa^{-1}
({e}/{m}) x$ the coordinate of $S^1$. A five-velocity
$u^A$ = $(u^a, u^5)$ can consequently be defined by
\be
u^A = \frac{d x^A}{d \sigma} . 
\label{3.8}
\ee
The components $u^a$ form the usual tangent-space four-velocity, which is related to
the spacetime four-velocity $u^\mu$ by $u^a=h^a{}_{\mu} u^\mu$, and $u^5$ is a strictly
internal component whose value will be determined by the unification process.

With the above definitions, and denoting by $\eta_{5 5}$ the fifth component of the
internal-space metric, if the conditions
$u^5 = - \kappa$
and
$\eta_{5 5} = - 1$
are satisfied, the action (\ref{3.1}) can be rewritten in the form
\begin{equation}
S=\int_{a}^{b}\left[- m \ c \ d\sigma - \frac{m}{c} {\mathcal A}^{A}{}_{\mu} \,
u^{B} \, \eta_{A B} \, dx^{\mu} \right] .
\label{3.11}
\end{equation}
Accordingly, the equation of motion (\ref{3.2}) becomes
\begin{equation}
c^2 h^{a}{}_{\rho} \frac{du_{a}}{ds} =
{\mathcal F}^{A}{}_{\rho \mu} \ u^{B} \ u^{\mu } \, \eta_{A B} .
\label{(3.13)}
\end{equation}
The trajectory of a charged particle submitted to both an
electromagnetic and a gravitational field, therefore, is
described by a Lorentz-type force equation. Furthermore, differently from curvature,
we see that torsion acts on
particles in the same way the electromagnetic field acts on charges, that is, as a force.

It is important to notice that, alternatively, we could have chosen
$u^5 = \kappa$
and
$\eta_{5 5} = 1$,
which would lead to the same action integral, and consequently to the same equation of
motion. As we will see, this choice corresponds to another metric convention for the
internal space. In principle, both conventions are possible. However, the unification
process will introduce a constraint according to which the choice of $\eta_{5 5}$ will
depend on the metric convention adopted for the tangent Minkowski space. 

In a gauge theory for the translation group, the gauge transformation is defined as a
local translation of the tangent-space coordinates,
\begin{equation}
\delta x^{a} = \delta\alpha^{b}P_{b}x^{a},
\end{equation}
with $P_{b} = \partial /\partial x^b$ the generators, and $\delta \alpha^{b}$ the
corresponding infinitesimal parameters.
In a unified teleparallel Kaluza-Klein model, a general gauge transformation is
represented by a translation of the five-dimensional internal space coordinates $x^A$,
\be
\delta x^A = \delta \alpha^B \, P_B \, x^A ,
\label{transin1}
\ee
where $P_B = {\partial}/{\partial x^B}$ are the group generators, and 
\[
\delta \alpha^B(x^\mu) \equiv \delta \alpha^B = \left(\delta \alpha^a, \delta \alpha^5
\right)
\]
are the transformation parameters. Analogously to the gauge potentials, we take
$\delta \alpha^5$ = $\kappa^{-1} ({e}/{m}) \delta \alpha$.
Furthermore, in the same way as in ordinary Kaluza-Klein models, we assume the
gauge potentials ${\mathcal A}^A{}_\mu$, and consequently the tetrad $h^a{}_\mu$ and
also the metric tensor $g_{\mu \nu}$, not to depend on the coordinate $x^5$.

The gauge gravitational field Lagrangian is~\cite{paper10}
\be
{\cal L}_G = \frac{h}{16\pi G}\left( \frac{1}{4} F^{a}{}_{\mu \nu}
F^{b}{}_{\theta \rho}g^{\mu \theta}N_{ab}{}^{\nu \rho}\right),
\label{2.36}
\ee
where $h = {\rm det}(h^{a}{}_{\mu})$, and due to the presence of a tetrad field,
we must have
\be
N_{ab}{}^{\nu \rho} = \eta_{ab} \; h_{c}{}^{\nu} \; h^{c \rho}
+ 2 \, h_{a}{}^{\rho} \; h_{b}{}^{\nu}
- 4 \, h_{a}{}^{\nu} \; h_{b}{}^{\rho}.
\label{2.38}
\ee
The unified Lagrangian, therefore, can be written in the form
\be
{\cal L} = \frac{h}{16 \pi G}\left( \frac{1}{4} {\cal F}^{A}{}_{\mu \nu}
{\cal F}^{B}{}_{\theta \rho}g^{\mu \theta}N_{AB}{}^{\nu \rho}\right) .
\label{4.25}
\ee
As no tetrad exists in the electromagnetic sector,
\be
N_{55}{}^{\nu \rho} = \eta_{55} h_{c}{}^{\nu} \; h^{c \rho} .
\ee
Thus, the Lagrangian (\ref{4.25}) becomes
\begin{equation}
{\cal L} = \frac{h c^4}{16 \pi G} \; S^{\rho \mu \nu} \; T_{\rho \mu \nu} 
+ \eta _{55} \, 
\frac{\kappa^{-2} e^2}{16 \pi G m^2} \left[ \frac{h}{4} 
F_{\mu \nu} F^{\mu \nu} \right] ,
\label{la1}
\end{equation}
where
\[
S^{\rho \mu \nu} = \textstyle{\frac{1}{2}} \left[ K^{\mu \nu \rho}
- g^{\rho \nu} \; T^{\theta \mu}{}_{\theta} + g^{\rho \mu} \;
T^{\theta \nu}{}_{\theta} \right] = - S^{\rho \nu \mu} ,
\]
with $K^{\mu \nu \rho}$ the contorsion tensor.

The first term of ${\cal L}$ is the gauge gravitational Lagrangian, which is equivalent
to the Hilbert-Einstein Lagrangian of general relativity.
In order the get Maxwell's Lagrangian from the second term, two conditions must be
satisfied. First, it is necessary that
\be
\kappa^2 = \frac{e^2}{16 \pi G m^2},
\ee
from where we see that $\kappa^2$ turns out to be proportional to the ratio between
electric and gravitational forces~\cite{narlikar}. Second, in order to have a
positive-definite energy for the electromagnetic field, and to get the appropriate
relative sign between the gravitational and electromagnetic Lagrangians, it is necessary
that
$\eta_{55}$ = $- 1$. With these conditions, the Lagrangian (\ref{la1}) becomes
\[
{\cal L} \equiv h \left( L_{G} + L_{EM} \right)
= \frac{h c^{4}}{16 \pi G} \; S^{\rho \mu \nu} \; T_{\rho \mu \nu} - 
\frac{h}{4} F_{\mu \nu} F^{\mu \nu} .
\]
That the Maxwell Lagrangian in four dimensions shows up from the Hilbert-Einstein
Lagrangian in five dimensions is usually considered as a miracle of the standard
Kaluza-Klein theory~\cite{paper15}. That the Hilbert-Einstein Lagrangian of general
relativity shows up from a Maxwell-type Lagrangian for a five-dimensional translation
gauge theory can be considered as the other face of the same miracle.

The functional variation of ${\cal L}$ in relation to $A^a{}_{\mu}$ leads to the
field equation
\be
\partial_\sigma (h S_{\lambda}{}^{\sigma \tau}) - 
\frac{4 \pi G}{c^4} (h t_{\lambda}{}^{\tau}) = 
\frac{4 \pi G}{c^4} \, (h \theta_{\lambda}{}^{\tau}) ,          \label{4.34}
\ee
where 
\be
t_{\lambda}{}^{\tau} =
\frac{c^{4}}{4 \pi G} \, \Gamma^{\mu}{}_{\nu \lambda} S_{\mu}{}^{\nu \tau}
+ \delta_\lambda{}^{\tau} L_G
\label{ptem} 
\ee
is the canonical energy-momentum (pseudo) tensor of the gravitational field, and
\be
\theta_{\lambda}{}^{\tau} \equiv h^{a}{}_{\lambda} \left[ 
-\frac{1}{h} \frac{\delta {\cal L}_{EM}}{\delta h^{a}{}_{\tau}} \right] = 
F_{\lambda \nu} F^{\tau \nu}
- \delta_{\lambda}{}^{\tau} L_{EM}
\ee
is the energy-momentum tensor of the electromagnetic field. On the other hand,
the functional variation of ${\cal L}$ in relation to $A_{\mu}$ yields the
teleparallel version of Maxwell's equation~\cite{vector}.

An interesting point that deserves to be mentioned is that since we have chosen
(\ref{1.1}) as the metric of the Minkowski space, the resulting metric of the
five-dimensional internal space will be
\be
\eta_{A B} = \mbox{diag}(+1, -1, -1, -1, -1) .
\label{m1}
\ee
This means that the fifth dimension must necessarily be spacelike, and the
metric with signature $(3,2)$ is excluded. On the other hand, if we had chosen
$\tilde{\eta}{}_{a b} = \mbox{diag}(-1, 1, 1, 1)$ for
the Minkowski space, it is easy to verify that we should have
$\eta_{55} = 1$. The resulting metric of the
five-dimensional internal space would be
\be
\tilde{\eta}{}_{A B} = \mbox{diag}(-1, +1, +1, +1, +1) ,
\label{m2} 
\ee
and the same conclusion would be obtained:
the fifth dimension must necessarily be spacelike, and the metric with signature
$(3,2)$ is excluded. The unification of the gravitational and electromagnetic Lagrangians,
therefore, imposes a constraint on the metric conventions for Minkowski and for the
electromagnetic internal manifold $S^1$. In fact, the choice between $\eta_{55}$ = $+1$ and
$\eta_{55}$ = $-1$ for the Killing metric of the $U(1)$ group depends on the metric
convention adopted for the Minkowski space. As a consequence, the metric of the
five-dimensional internal space turns out to be restricted to either Eq. (\ref{m1}) or
(\ref{m2}). Metrics with signature $(3,2)$, which would imply a time-like fifth dimension,
are excluded.

\section{Matter Fields}

Let us consider now a matter field $\Psi$. In contrast to the gauge fields, it
depends on the coordinate $x^5$:
\[
\Psi = \Psi(x^\mu, x^5) \; .
\]
Under a generalized gauge translation, it behaves as
\be
\delta \Psi = \delta \alpha^A \, P_A \, \Psi .
\label{psigauge}
\ee
Its covariant derivative, consequently, is
\be
D_\mu \Psi = \partial_\mu \Psi + c^{-2} {\mathcal A}^A{}_\mu P_A \, \Psi,
\ee
with the gauge transformation of ${\mathcal A}^A{}_\mu$ given by
\be
\delta {\mathcal A}^{B}{}_\rho = -
c^{2} \partial_\rho \delta \alpha^B .
\label{agaugein}
\ee
Separating the gravitational and electromagnetic components, we obtain
\be
D_\mu \Psi = \partial_\mu \Psi + c^{-2} A^a{}_\mu P_a \Psi + 
\kappa^{-1} \frac{e}{m c^{2}} A_\mu \, P_5 \, \Psi \; .
\label{psicova}
\ee

Now, as $x^5$ is the coordinate of the internal manifold $S^1$, we assume 
\be
\Psi(x^\mu, x^5) = \exp \left[ i \, \kappa \, \frac{2 \pi}{\lambda_C} \, x^5 \right] \,
\psi(x^\mu) \; ,
\ee
with $\lambda_C = ({h}/{m c})$ the Compton wavelength of the particle under
consideration. Consequently, a translation in the coordinate $x^5$ turns out to be
a $U(1)$ gauge transformation, and a translation in the coordinates $x^a$
turns out to be a gauge transformation related to the translation group $T^4$.
For a simultaneous translation in the five coordinates $x^A$, we see from
Eq. (\ref{psigauge}) that
\be
\delta \Psi = \delta \alpha^a \, \partial_a \Psi + 
\delta \alpha \, \left( \frac{i e c}{\hbar} \right) \Psi .
\ee
The corresponding minimal couplings are given by the covariant derivative
\be
D_\mu \Psi = h^a{}_\mu \, \partial_a \Psi +
\frac{i e}{\hbar c} \, A_\mu \, \Psi .
\label{fpsicova}
\ee
By defining $A_a=h_a{}^\mu \, A_\mu$, we can rewrite Eq. (\ref{fpsicova}) in the form
\be
D_\mu \Psi = h^a{}_\mu \, D_a \Psi ,
\ee
with $D_a \Psi$ the electromagnetic covariant derivative in Min\-kows\-ki space.
As usual, the com\-mu\-ta\-tor of co\-va\-riant derivatives yields the
field strength:  
\[
[D_\mu,D_\nu]\Psi = c^{-2} {\mathcal F}^A{}_{\mu \nu} P_A \Psi =
c^{-2} F^a{}_{\mu \nu} P_a \Psi + 
\frac{ie}{\hbar c} F_{\mu \nu} \Psi .
\]

\section{Conclusions}

By replacing the general relativity par\-a\-digm with a gauge par\-a\-digm, a
five-di\-men\-sional Maxwell-type trans\-la\-tional gauge theory on a four-di\-men\-sional
spacetime has been constructed. In this theory, gravitation is at\-trib\-ut\-ed to torsion, and
the electromagnetic field str\-ength appears as the fifth gauge-component of the torsion
tensor. Because of the fact that torsion, like the electromagnetic field, plays the role of a
force, the unification in this approach seems to be much more natural than in ordinary
Kaluza-Klein theories. 

An important feature of this model is that no scalar field is generated by the unification
process.  Accordingly, no unphysical constraints appear, and the gravitational action can
naturally be truncated at the zero mode.  Furthermore, the cylindric condition can also be
naturally imposed for matter fields, which corresponds to keep only the $n=1$ mode in their
Fourier expansions.  Consequently, the infinite spectrum of massive new particles is
eliminated, strongly reducing the redundancy present in ordinary Kaluza-Klein theories. A
similar achievement has been obtained recently by a modified Kaluza-Klein theory in which
the internal coordinates are replaced by generators of a noncommutative
algebra~\cite{madore}. In this model, no truncation to eliminate extraneous modes is
necessary as only a finite number of modes is present.  Concerning this point, the
teleparallel version of Kaluza-Klein seems to be much more appropriate to deal with models
involving noncommutative geometry.  In fact, as the spacetime underlying such a model is a
spacetime presenting torsion, but no curvature, the usual difficulty for defining
curvature~\cite{curve}, and consequently for choosing the action functional~\cite{connes},
is avoided.

The generalization to non-Abelian gauge theories a\-mounts to introduce a $4+N$ dimensional
internal space, formed by the product between the Minkowski space and a compact Riemannian
manifold. As in the electromagnetic case, the teleparallel unification turns out to be much
more natural than in ordinary Kaluza-Klein models since both gravitational and Yang-Mills
fields are described by a gauge theory, the Yang-Mills field strength appearing as extra
gauge-components of torsion. In other words, the gravitational and the gauge
field strengths are different components of a unique tensor. Another important point
concerns the relation between geometry and gauge theories. According to ordinary
Kaluza-Klein models, gauge theories emerge from higher-dimensional geometric theories as a
consequence of the dimensional reduction process. According to our approach, however,
gauge theories are the natural structures to be introduced, the four-dimensional geometry
(gravitation) emerging from the noncompact sector of the internal space. In fact, only
this sector can give rise to a tetrad field, which is responsible for the geometrical
structure (either metric or teleparallel) induced in spacetime. Furthermore, as the gauge
theories are introduced in their original form --- they do not come from geometry --- the
unification, though not trivial, turns out to be much more natural and easier to be
performed.

Finally, it should be mentioned that several different Kaluza-Klein models in spacetimes
with torsion have already been considered.  However, most of them are constructed on a
five-dimensional spacetime, and are consequently completely different from the model presented
here~\cite{kkt}.

\section*{Acknowledgments}  

The authors would like to thank FAPESP-Brazil, CAPES-Brazil and CNPq-Brazil for financial
support.

\end{document}